\documentclass[aps,pre,preprint,onecolumn,showpacs,amsmath,amssymb]{revtex4-1}

\usepackage{graphicx}
\usepackage{bm}
\usepackage{color}
\usepackage{soul} 
\usepackage{lineno}
\usepackage{textgreek}
\usepackage{caption}
\usepackage{array}

\begin{document}

%\title{Effect of dispersants on interfacial attachment and growth of bacteria}
%\title{Effect of dispersants on attachment and growth of bacteria to microscale oil droplets}
%\title{Microscale attachment and growth of bacteria to dispersant-decorated oil droplets}
%\title{Microfluidic observations of attachment and growth of bacteria on dispersant-decorated oil droplets}
%\title{Direct observations of attachment and growth of bacteria on oil droplets}

%\title{Effect of dispersants on the colonization of oil droplets by bacteria using microfluidics}

\title{Effect of dispersants on bacterial colonization of oil droplets: a microfluidic approach}

\author{Vincent Hickl$^{a}$}

\author{Gabriel Juarez$^{b}$}
\thanks{Email address}\email{gjuarez@illinois.edu}

\affiliation{$^{a}$Department of Physics, University of Illinois at Urbana-Champaign, Urbana, Illinois, 61801, USA}

\affiliation{$^{b}$Department of Mechanical Science and Engineering, University of 
Illinois at Urbana-Champaign, Urbana, Illinois 61801, USA}

\date{\today}

%\linenumbers

%--------------%
%%% Abstract %%%
%--------------%
\begin{abstract}

Bacteria biodegradation of immiscible oil requires cell-droplet encounters, surface attachment, and hydrocarbon metabolism. Chemical dispersants are applied to oil spills to reduce the mean dispersed droplet size, thereby increasing the available surface area for attachment, in attempts to facilitate bacterial biodegradation. However, their effectiveness remains contentious as studies have shown that dispersants can inhibit, enhance, or have no effect on biodegradation. Therefore, questions remain on whether dispersants affect surface attachment or cell viability. Here, using microfluidics and time-lapse microscopy, we directly observe the attachment and growth of the marine bacterium, \emph{Alcanivorax borkumensis}, on stationary crude oil droplets ($5$ \textmu m $< R < 150$ \textmu m) in the presence of Corexit 9500. We show that the average colonization time, or the time comprised of encounters, attachment, and growth, is dependent on droplet size and primarily driven by diffusive encounters. Our results suggest that dispersants do not inhibit or enhance these biophysical processes.

\end{abstract}

%\pacs{XX.xx.xx}

\maketitle

%------------------%
%%% Introduction %%%
%------------------%
\section{Introduction}

Oil spills remain a common and extremely dangerous threat to marine ecosystems around the world. Understanding the fate of spilled oil in the short and long term is crucial for designing appropriate oil spill response techniques to remove the oil and protect marine life. Spilled oil often takes the form of microscopic droplets which can remain suspended in the water column for months \cite{Camilli2010,Passow2021}. Over these long time scales, oil can be transported over hundreds of kilometers \cite{North2015,French-McCay2019}, making an accurate assessment of its fate very difficult. Some estimates have suggested that up to 70\% of the spilled oil from the Deepwater Horizon (DWH) disaster may have formed a deep-sea plume \cite{Ryerson2012,Passow2021} of oil, which contained droplets ranging from $\leq 1$ \textmu m to hundreds of \textmu m in radius \cite{Li2015}. The fate of these microscopic droplets has important implications for the fate of marine ecosystems on scales many orders of magnitude larger than the droplet size.

%Estimates of the fate of spilled oil are very uncertain, as a full mechanistic model of biodegradation remains elusive and accurate in-vivo assessments are impractical on such large scales. 

Marine bacteria play an important role in oil spill remediation, as they can use the hydrocarbons in crude oil as a nutrient source, and ultimately remove some of the oil from their environment. More than 30\% of the spilled oil from DWH may have eventually been biodegraded by bacteria and other organisms \cite{Passow2016}. Oil-degrading bacteria typically constitute only a small portion ($<1$\%) of marine microbial populations, but their numbers increase dramatically when encountering the high concentrations of oil produced by a spill \cite{MartinsdosSantos2010,Cappello2007,Harayama1999}. One ideal model organism is \textit{Alcanivorax borkumensis}, an oil-degrading strain which has been studied extensively due to its high prevalence in oil-contaminated waters in the Gulf of Mexico and elsewhere \cite{Head2006,Schneiker2006,Kimes2014}. To access the immiscible hydrocarbons in the oil, bacteria must attach to the surface of suspended oil droplets. Thus, to better understand biodegradation mechanics, one must study the attachment and subsequent growth of bacteria at the oil droplet surface. 

Previous work has suggested that the presence of chemical dispersants can have significant effects on the biodegradation of oil droplets by bacteria. The introduction of dispersants into oil-contaminated water is a common oil-spill response designed to reduce the mean radius of oil droplets, increasing the surface-area-to-volume ratio and thus the bio-availability of the oil to marine organisms. Perhaps the most well-studied chemical dispersant is Corexit 9500, which was used extensively following the DWH spill. Batch reactor experiments involving crude oil, bacteria, and dispersants have yielded contradictory results, with different studies reporting enhancement \cite{Prince2013,McFarlin2014}, no effect \cite{Macnaughton2003,Swannell1999}, and inhibition \cite{Kleindienst2015a, Rahsepar2016} of biodegradation due to dispersants. 

As a result, new research in recent years has increasingly focused on the droplet scale to directly assess the effect of dispersants on bacterial attachment to oil droplets. Studies have revealed a multitude of interactions between Corexit 9500 and oil-degrading bacteria. While dispersants increase the area available for colonization, they also reduce the surface tension, lowering the attachment energy \cite{Dewangan2018}. It has been noted that dispersants can be toxic to cells, for example by disrupting the cell membrane \cite{Kleindienst2015B}. On the other hand, some experimental results suggest that the oil-degrading bacterium \textit{A. borkumensis} can metabolize some components of Corexit 9500 \cite{Abbasi2018}. Additionally, different dispersants can have drastically different effects on cell attachment depending on whether they are anionic, nonionic, or cationic, particularly when the bacteria of interest produce their own biosurfactants \cite{Bookstaver2015}. These complex interactions highlight the need for well-designed experimental protocols which can distinguish between the various effect of dispersants on biodegradation and reveal the precise interfacial physics that ultimately determine whether bacteria can effectively attach to, grow on, and colonize the droplet surface.

Here, we present experimental results and simulation data on the colonization of microscopic oil droplets with and without chemical dispersants. We created a custom microfluidic device allowing for the continuous and simultaneous observation of hundreds oil droplets of radii $1$ \textmu m $\leq R\leq150$ \textmu m over several days. Bacteria, dispersants, and nutrients were injected into the device to identify the effects of droplet size and dispersant concentration on bacterial attachment and growth at the droplet surface. Prolonged cell growth at the interface leads to buckling of the oil droplets, and the buckling time is a measure of how long it takes for cells to form a monolayer at the droplet surface. Detailed analysis of buckling times, combined with simulation data and experiments at flat oil-water interfaces, shows that chemical dispersants do not significantly affect the initial attachment or growth of cells at the droplet surface. Using microfluidics and image analysis for direct droplet observation thus allows us to precisely distinguish between different physical effects of chemical dispersants on oil-degrading bacteria at the surface of oil droplets.

\section{Experimental methods}

The bacterial strain used in this study was \emph{A. borkumensis} (ATCC 700651), a rod-shaped, non-motile, alkane-degrading marine bacterium. Bacteria were grown by incubating cells in 5 mL of culture medium for 24 hours at $30 ^{\circ}$C in an orbital shaker at $180$ rpm. The culture medium consisted of 37.4 g/L 2216 marine broth (BD Difco) and 10 g/L sodium pyruvate. The average doubling time was measured to be 1.6 hours using a Biowave CO8000 cell density meter. Prior to the start of each experiment, a cell culture in the late exponential phase was diluted using culture medium to an optical density of 0.18. This suspension was then diluted $50\times$ in artificial seawater (ASW), which was then injected into the microfluidic device described below. Using a hemocytometer, the cell concentration of this final suspension was measured to be $2.5\times10^7$ cells/mL.

Experiments were performed in custom microfluidic chambers sealed with silicone rubber sheets (McMaster-Carr 3788T22) and epoxy. The chamber dimensions were 3 cm in length, 1 cm in width, and 1.5 mm in height. Inside the chamber, stationary oil droplets were created by depositing $10$ \textmu L of unweathered MC252 crude oil on a glass microscope slide, forming a thin film. MC252 is a light sweet crude oil having a density of $0.83$ g/mL, an interfacial tension of $20$ mN/m, and a viscosity of $3.9$ mPa s at $32^{\circ}$C \cite{Daling2014}.

Then, $450$ \textmu L of ASW was dripped on top of the oil film. The addition of water initiated the formation of microscopic oil droplets in the form of spherical caps, wetted to the glass slide, and having a size range of $1$ \textmu m $< R < 200$ \textmu m (See Supplemental Material). On average, the caps contained $93\pm3\%$ of the volume of a sphere of the same diameter. Lastly, before the chamber was sealed, polyethylene tubing was attached to serve as inlets and outlets. 

Next, a syringe pump was used to inject 2 mL of a solution of chemical dispersant Corexit 9500 into the device. Corexit concentrations used were 0 ppm (control) and 100 ppm (protocol 1). The volume injected was roughly twice that of the chamber volume, and the injection was performed gradually over 20 minutes to ensure that Corexit was present throughout the entire microfluidic device and decorated the surface of all the oil droplets. The chamber was then flushed with ASW (1 mL injected over 10 minutes), to mimic the rapid dilution of dispersants likely to occur at sea.

Then, 2 mL of a bacteria suspension of \textit{A. borkumensis} in ASW were injected into the device over 20 minutes. This step allowed a small number of cells to attach to the oil droplets, where they would continue to grow and eventually form a monolayer at the droplet surface. To maintain cultures within the chamber for at least 24 hours, a solution of diluted culture medium and ASW (1:9) was injected slowly into the device (flow rate of $0.01$ mL/min) for the remained of the experiments, which lasted between 24 and 36 hours. In a separate set of experiments (protocol 2), this diluted solution of culture medium was supplemented with Corexit 9500 at a concentration of 100 ppm (protocol 2). In protocol 2, no Corexit was injected before the addition of bacteria. The different experimental protocols are shown schematically in Fig. 1.

In each experiment, at least $100$ droplets were monitored using time-lapse phase contrast and epifluorescence microscopy for 24 hours. Using an automated motor-controlled stage, images of the midplane of each droplet were captured every 10 minutes at $40\times$ magnification with a sCMOS camera (Andor Zyla 4.2). All experiments were performed at a room temperature of $20 ^\circ$C.

To observe bacteria growing at the oil-water interface, a 100-mesh SEM grid (205-\textmu m aperture) was placed onto the thin film of oil before adding ASW into the microfluidic device. Oil thus became pinned to the edges of the grid, forming flat interfaces at the bottom of the microfluidic chamber. The division of cells adsorbed to the interface could be seen clearly using timelapse microscopy (1 image every 5 minutes), and measurements of the doubling time were made. A custom MATLAB algorithm was also used to analyze the cell growth and division.

\begin{figure*}
\centering
    \includegraphics[width=0.9\linewidth]{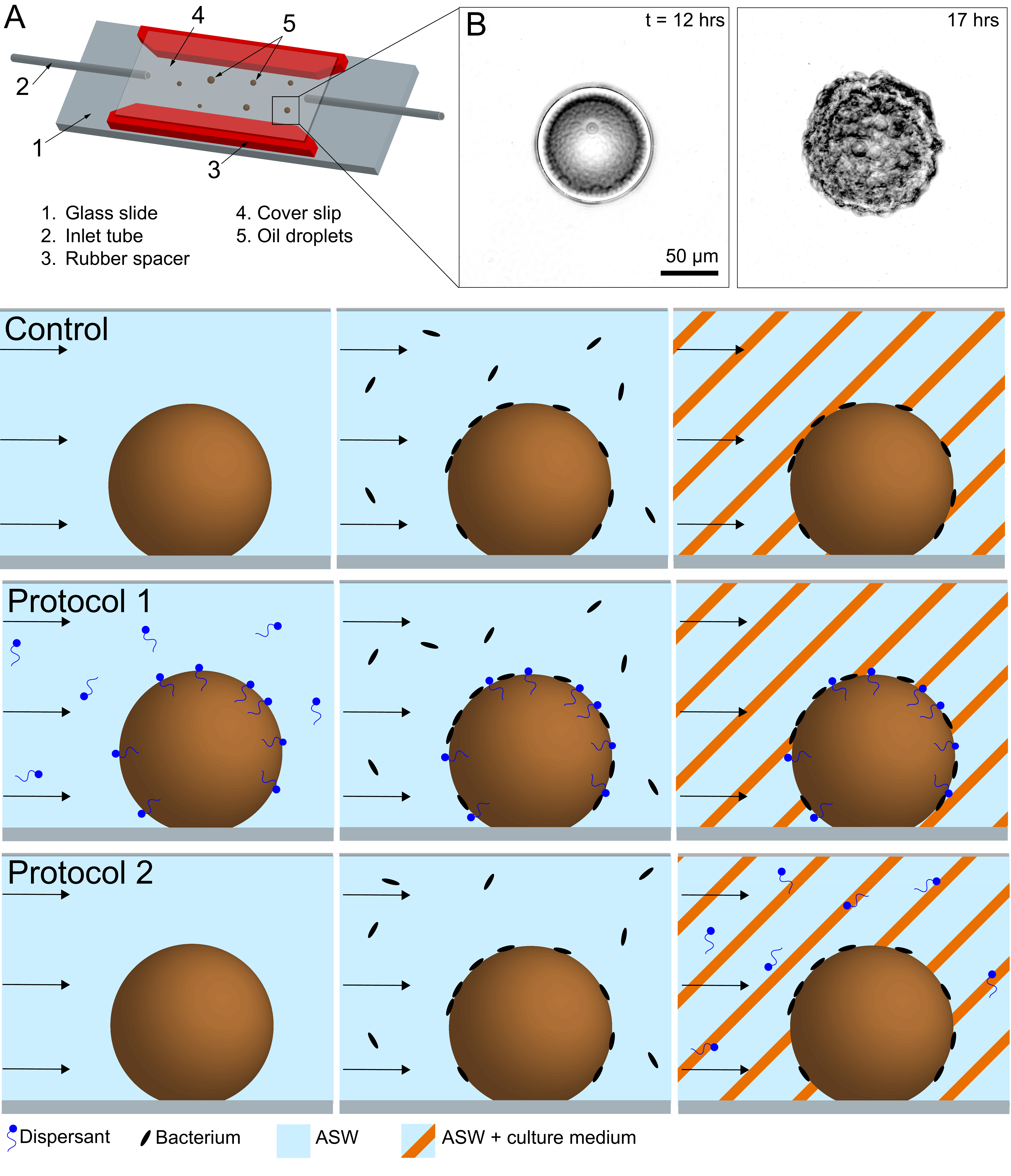}
\caption{
(A) Schematic of microfluidic device. Droplets are pinned to a glass slide and enclosed in a chamber into which dispersants, cells, and nutrients are injected. (B) Sample droplet with $R=51$ \textmu m before ($t = 12$ h) and after buckling (17 h). (Control) Cells are injected into the device, then flushed out before nutrients diluted in ASW are added continuously. (Protocol 1) Dispersants decorate droplets, then cells and later nutrients are injected. (Protocol 2) Cells are injected, then dispersants are injected with nutrients in ASW for the remainder of the experiment.
}
    \label{fig:fig1}
\end{figure*}

\section{Results}

The growth and divison of \textit{A. borkumensis} confined to the surface of oil droplets leads to interfacial deformations. In our experiments, crude oil droplets are pinned to a glass microscope slide at the bottom of a microfluidic device. At the beginning of each experiment, cells diffusively encounter droplets where they adsorb to the surface and subsequently grow. The energy of attachment is estimated to be on the order of $10^6\times kT$, meaning cell attachment minimizes the energy of the system, and bacteria cannot spontaneously detach from droplets due to thermal fluctuations. Over the course of 10-20 hours, cells metabolize hydrocarbons, grow, divide, and form a monolayer on the droplet surface. Because of the finite size of the droplet, stress at the surface increases due to continued growth in the monolayer, which overcomes surface tension and buckles the droplet surface (see supplementary video). Thus, the buckling time, measured from the start of the experiment, is an estimate of how long it takes for bacteria to fully colonize an oil droplet. The buckling time is a function of the droplet radius, the encounter rate between cells and the droplet, the proportion of cell encounters that lead to attachment, and on the growth rate of cells at the surface.

Buckling time depends on droplet size. Kymographs of two droplets with radii $14 \text{ \textmu m } \leq R\leq60$ \textmu m, respectively, are shown in Figure 2A. The buckling time can be identified clearly as the location in the graph where the surface of the droplet becomes disturbed and its apparent size begins to increase, as shown by the dotted lines in Fig. 2A. The buckling times for these particular droplets were 14.7 h and 19.3 h, respectively. 

Average buckling time increases with droplet radius. Measurements of the buckling time for 253 droplets with radii $4.3$ \textmu m $\leq R\leq139$ \textmu m are shown in Figure 2. For droplets with $R\lesssim40$ \textmu m, there is a wide variation in buckling times. In fact, both the longest and shortest buckling times, 10.7 h and 21.6 h, respectively, are observed in this size range. For droplets with $R\gtrsim40$ \textmu m, buckling times fall within a narrower range, and buckling times appear to approach an asymptotic value of about $17$ h. The mean buckling time of all droplets is $16.0\pm0.1$ h.

\begin{figure*}[h]
\centering
    \includegraphics[width=\linewidth]{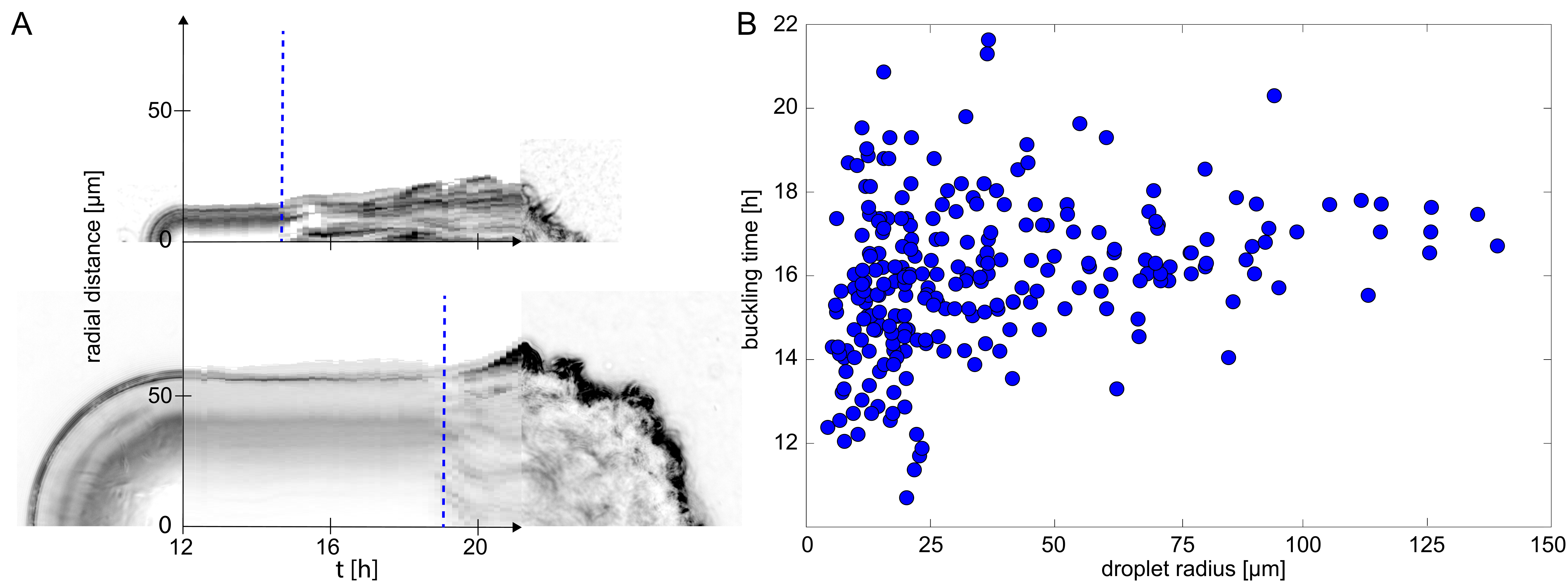}
\caption{
(A) Kymographs of two droplets ($R=14$ \textmu m and $R=60$ \textmu m, respectively) which buckle at different times ($T_b=14.7$ h and $T_b=19.3$ h). (B) Plot of measured buckling times as a function of droplet radius for $N=253$ droplets from control experiments. 
}
    \label{fig:fig2}
\end{figure*}

\subsection{Simulations of interfacial attachment and growth}

Simulations of bacteria encountering, attaching to and growing on oil droplets validated the relationship between droplet size and buckling time in the absence of dispersants. To simulate attachment we consider diffusive encounters and exponential cell growth as described in more detail below. The mean flux of bacteria diffusing to a spherical droplet of radius $R$ can be estimated as:
\begin{equation}
    Q=4\pi DRC_\infty
\end{equation}
where $D$ is the diffusion coefficient of the cells and $C_\infty$ is the background cell concentration \cite{Kiorboe2008}. The total number of cells reaching the droplet surface in a time interval $T$ is $N_0=QT$. In our experiments, cells can encounter droplets from the time the injection of the cell suspension starts, to the time cells are flushed out by the injection of another fluid (nutrients diluted in ASW) begins. This time interval gives us the value of $T$ used in simulations. We suppose the cells divide with a constant doubling time. Then, the buckling time is determined by the time interval required for cells to form a monolayer completely covering the droplet. That is, 
\begin{equation}
    2^{t/T_d}N_0A_b=4\pi R^2
\end{equation}
where $A_b$ is the contact area of a single cell on the droplet surface. Using this approach, the buckling time is deterministic for each droplet, and monotonically increases with $R$, as shown by the blue circles of Fig. 3. While this deterministic model reproduces the overall positive correlation between buckling time and droplet radius, it fails to capture the stochastic nature of cell attachment in lab experiments or marine environments.

\begin{figure*}
\centering
    \includegraphics[width=\linewidth]{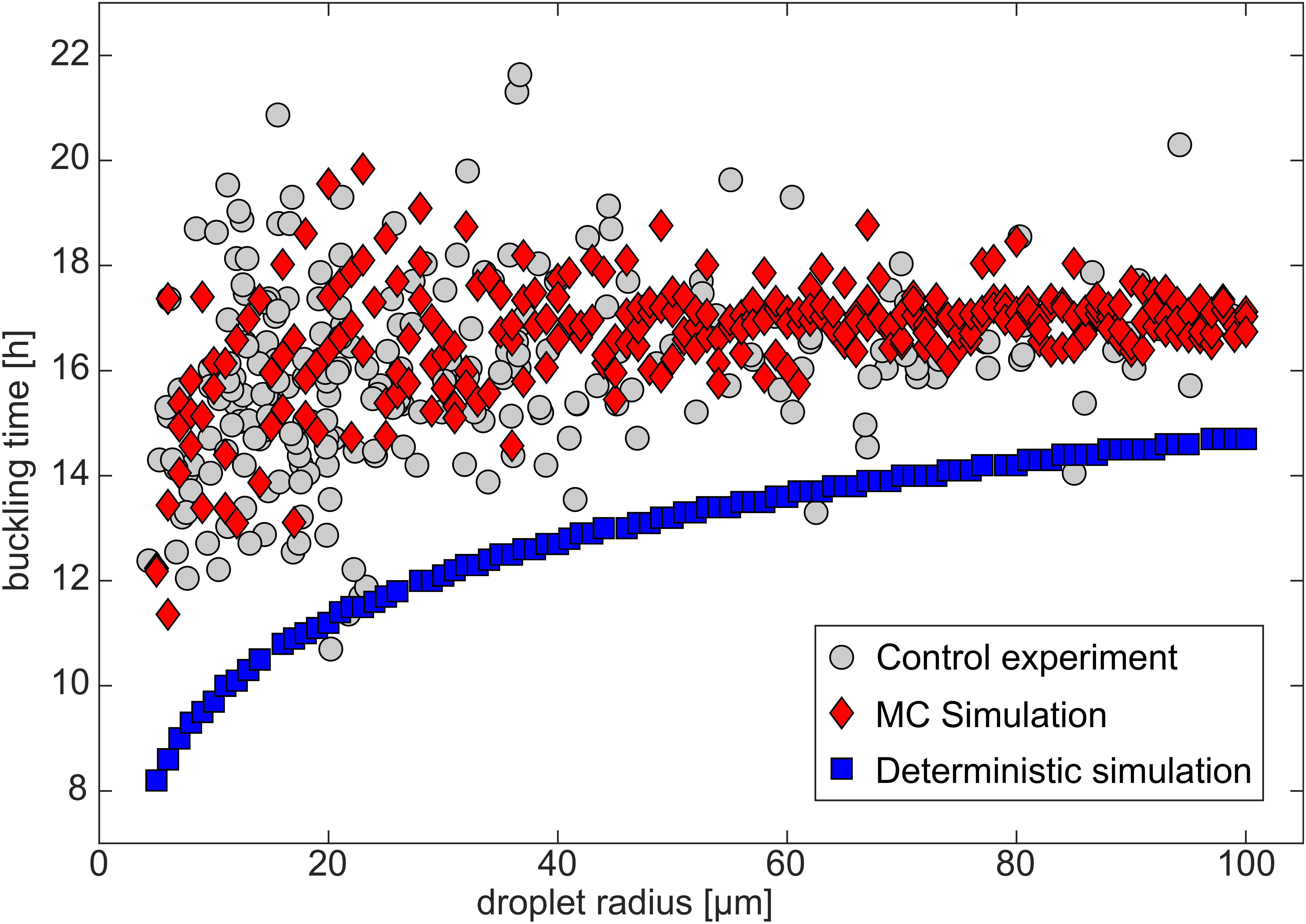}
\caption{Buckling times as a function of droplet radius for deterministic model of attachment and growth (blue), Monte Carlo simulations of attachment (red), and compared to control experiments (gray). Except in deterministic case, small droplets can buckle later than larger ones.
}
    \label{fig:fig3}
\end{figure*}

The precise dependence of buckling time on droplet radius observed in experiments was recovered in agent-based Monte Carlo simulations that implement single-cell diffusion. Bacteria positions were initialized at random positions in a volume of water surrounding an oil droplet of radius $R$. Cell positions evolved diffusely at each time step:
\begin{equation}
    x_{i,t+\Delta t}=x_{i,t}+\sqrt{D\Delta t}N_i
\end{equation}
where $i\in\{1,2,3\}$ in three dimensions, $\Delta t$ is the duration of each time step, and $N_i$ is a standard normal variable generated separately for each dimension. After the time interval $T$ the number of cells that have encountered each droplet is recorded, and the buckling time is calculated as before. This process was iterated 288 times for droplets with $5$ \textmu m $\leq R\leq100$ \textmu m. The dependence of buckling time on droplet size in simulation matched the one observed in experiments, as shown in Fig. 3. In Monte Carlo simulations, as in experiments, there is considerably more variation in the buckling time for droplets with $R \lesssim 40$ \textmu m. In both cases, buckling times are more uniform for $R \gtrsim 40$ \textmu m, with a mean buckling time of $16.6\pm0.1$ h and $16.97\pm0.04$ h for experiments and simulations, respectively. The discrete and stochastic nature of cell diffusion leads to some small droplets having greater buckling times due to the low encounter rate. In other words, it can take a long time for the first cell to attach to a droplet and start dividing on the droplet surface.

\subsection{Effect of Corexit 9500 on buckling times}

Having characterized and validated the dependence of buckling times on droplet size, we now turn to the effect of dispersants on the attachment and growth of bacteria at the droplet surface. Since buckling requires both these factors, a change in the buckling time suggests that either cell attachment or cell growth rate may be affected. To distinguish between the effects of Corexit on attachment and growth, our microfluidic device was designed such that bacteria, dispersants, and nutrients can be introduced in different orders. In protocol 1, dispersants are introduced first, then cells, then finally a nutrient solution. In protocol 2, cells are introduced first, and a solution containing nutrients and dispersants is injected until buckling occurs, as shown in Fig. 1.

To test the effect of dispersants on cell attachment to oil droplets, a high concentration (100 ppm) of Corexit 9500 was injected to the microfluidic device after droplet formation, decorating the droplet surface (protocol 1). The fluid in the device was then diluted with ASW to replicate realistic marine conditions, where high concentrations of dispersant cannot persist in the bulk. Bacteria were then introduced into the device, leading to some initial cell attachment to the droplet surface. Finally, a dilute solution of growth medium in ASW was injected for the remainder of the experiment to promote cell growth. Thus, these experiments can be used to test the attachment of bacteria to the droplet surface decorated with dispersant molecules. If Corexit inhibits cell attachment, buckling times in experiments using protocol 1 will be longer compared to the control. If Corexit only affects growth rates, buckling times will remain the same. 

Chemical dispersant Corexit 9500 increases buckling times by less than 10\% when it is introduced prior to cell attachment. The buckling times for $159$ droplets, as well as the control experiment with no dispersant, are shown in Fig. 4a. The mean buckling time with 100 ppm Corexit is $17.2\pm0.1$ h compared to $16.0\pm0.1$ h for the control. The overall distribution of buckling times as a function of droplet size is very similar in both cases. While the difference in buckling times is statistically significant, it is much less than the variation in buckling times in each population. This results suggests that the prior presence of dispersants at the droplet surface slightly reduces cell attachment, but ultimately does not significantly inhibit the ability of bacteria to grow on and deform the droplets.

\begin{figure*}
\centering
    \includegraphics[width=\linewidth]{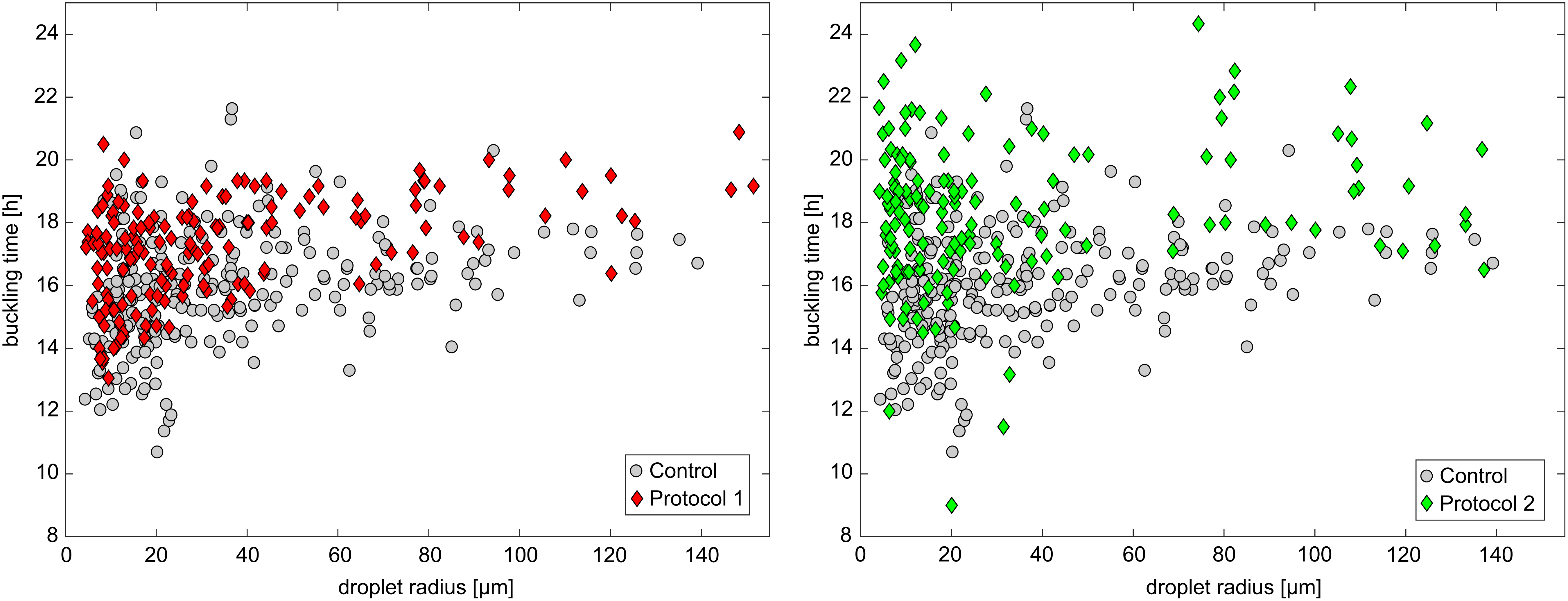}
\caption{
(Left) Buckling times as a function of droplet radius for protocol 1: 100 ppm of Corexit is injected before cells (red), compared to control (grey). (Right) Buckling times as a function of droplet radius for protocol 2: 100 ppm of Corexit is injected after cells (green) for the remainder of the experiment, compared to control (grey).
}
    \label{fig:fig4}
\end{figure*}

To test the effect of dispersants on interfacial cell growth, a separate set of experiments was conducted (protocol 2), in which bacteria were introduced to the device before any dispersants. Once cells had attached to the droplet surface, the dilute solution of growth medium was supplemented with 100 ppm Corexit, and injected for the remainder of the experiment. Thus, these experiments would show whether the presence of dispersants in the bulk could affect the growth of cells at the surface leading to buckling. If Corexit inhibits (enhances) cell growth rates, buckling times in experiments using protocol 2 will be longer (shorter) compared to the control. 

Corexit 9500 slightly increases buckling times when introduced after the bacteria. The buckling times for $146$ such droplets, as well as the control experiment with no dispersants,  are shown in Fig. 4b. The mean buckling time with dispersants is $18.3\pm0.2$ h. As before, there is considerable overlap between the buckling times of these experiments with the control data. the distributions of buckling times as a function of droplet size is qualitatively the same. This result suggests that there may be a reduction in the bacteria's growth rate when dispersants are present in the bulk. 

The increase in buckling times compared to control experiments is statistically significant ($p\leq0.05$) for nearly all droplet radii with both experimental protocol 1 and 2, as shown in Table I. Thus, the effect of dispersants on buckling does not depend on droplet size. Since buckling times increase regardless of when dispersants are introduced, this result cannot be explained by a decrease in cell attachment alone or by a slower cell growth rate alone. 

\begin{table}[!h]
\begin{center}
\begin{tabular}{ |>{\centering\arraybackslash}m{9em}||>{\centering\arraybackslash}m{6em}|>{\centering\arraybackslash}m{6em}|>{\centering\arraybackslash}m{6em}|>{\centering\arraybackslash}m{6em}| } 
 \hline
 Droplet radius [\textmu m] & Monte Carlo & Control & Protocol 1 & Protocol 2 \\ 
 \hline \hline
 0--20 & $15.4\pm0.5$ & $15.5\pm0.4$ & $16.6\pm0.4$* & $18.2\pm0.5$* \\ 
 \hline
 20--40 & $16.8\pm0.3$* & $16.0\pm0.4$ & $16.8\pm0.4$* & $17\pm1$ \\ 
 \hline
 40--60 & $16.9\pm0.2$ & $16.5\pm0.5$ & $18.1\pm0.5$* & $19\pm1$* \\ 
 \hline
 60--80 & $17.0\pm0.1$* & $16.3\pm0.5$ & $18.1\pm0.6$* & $20\pm2$* \\ 
 \hline
 80--100 & $17.0\pm0.1$ & $16.8\pm0.7$ & $18.8\pm0.9$* & $20\pm2$* \\ 
 \hline
 100--120 &  & $17.2\pm0.8$ & $19\pm1$* & $19\pm1$* \\ 
 \hline
 120--140 &  & $17.1\pm0.4$ & $18\pm1$ & $19\pm1$* \\ 
 \hline
\end{tabular}
\caption{Mean buckling times [h] for different droplet size ranges in each experimental protocol. Each cell shows the 95\% confidence interval. Asterisks denote statistically significant differences from the control experiment ($p\leq0.05$).}
\end{center}
\end{table}

\subsection{Direct visualization of cell growth}

To examine the effect of Corexit on interfacial cell growth through direct visualization, experiments were conducted on a flat oil-water interface, such that cell division could be observed and doubling times could be measured. The oil-water interface was pinned by the edges of an SEM grid to ensure long-term stability. Bacteria were introduced at a low concentration such that only $1-10$ cells were initially adsorbed to each grid square, as shown in Fig. 5 (left). Then, growth medium at the same concentration as in protocols 1 and  2 was injected to promote interfacial growth. As cells divide, they remain adsorbed to the interface, eventually forming a monolayer, as shown in Fig. 5 (middle). Their doubling times were measured using the frames in the timelapse when each cell splits in two, as shown in the inset of Fig 5. 

\begin{figure*}
\centering
    \includegraphics[width=\linewidth]{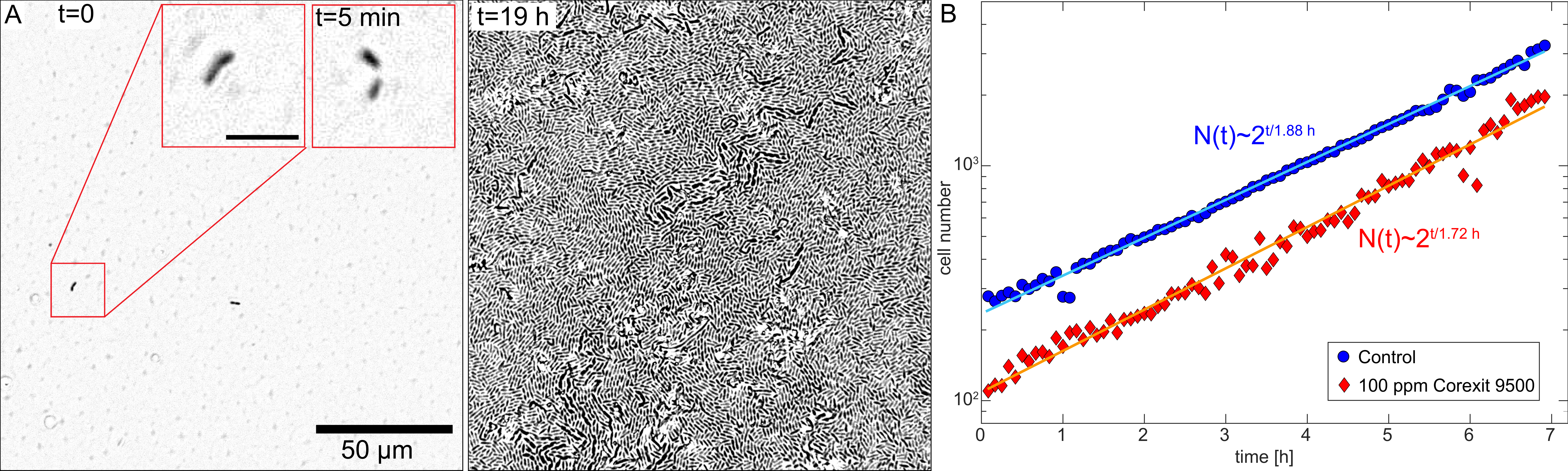}
\caption{
(A) t=0: Cells beginning to grow at a flat oil-water interface. Inset: two consecutive frames (5 mins apart) showing a cell dividing. The interval between successive divisions is the cell's doubling time. Inset scale bar is $10$ \textmu m. t=19 h: Cell monolayer at the flat interface after 19 h of continuous growth. (Right) Sample plots of cell number on the interface over time with and without Corexit. Only the period of exponential growth is shown. Lines show the best fit, which give an estimate of the doubling time in both experiments. 
}
    \label{fig:fig5}
\end{figure*}

Doubling times of \textit{A. borkumensis} at the oil-water interface are unaffected by Corexit 9500. Experiments were conducted in duplicate both without dispersants and with Corexit 9500 at a concentration of 100 ppm. The mean doubling times of individual cells is $1.74\pm0.08$ h in control experiments ($N=74$) and $T_d=1.69\pm0.03$ h with 100 ppm Corexit ($N=50$). Thus, the presence of Corexit in the bulk has no significant effect on the growth rate. Doubling times for the growing monolayer were also measured by counting the total number of cells $N$ at the interface using a custom image analysis algorithm. The increase in $N$ over time was exponential for several hours, as shown in figure 5 (right). Fitting these curves gives doubling times of $1.88\pm0.02$ h and $1.72\pm0.02$ h without and with Corexit, respectively. This result suggests that the increase in buckling times shown in Fig. 4b was not caused by a reduction in the rate of cell growth or division. Notably, we observed no detachment of cells from the interface either in the absence or presence of dispersants.

\section{Discussion}

Using a novel microfluidic device, we showed that the initial colonization of crude oil droplets by oil-degrading bacteria depends on the droplet size $R$, but is not strongly affected by chemical dispersant Corexit 9500 when the distribution of droplet radii is the same across experiments. Through direct observation of hundreds of individual droplets with different radii, our approach can provide new insights into the role of dispersants in oil biodegradation, which remains a contentious topic in the literature. 

Bacterial biodegradation of immiscible oil occurs at the droplet surface, meaning the degradation rate can be limited by the finite area of the surface. Buckling time is a useful way to quantify the colonization of droplets because it corresponds to the time when bacteria have covered the droplet surface in a monolayer, and because buckling increases the droplets' surface area, which leads to higher degradation rates. Our control experiments without dispersants revealed the dependence of buckling times on droplet radius. Small droplets tend to be colonized more quickly, but the variation in buckling is also much larger for small droplets. Thus, both the longest and shortest buckling times were observed on droplets with $R\lesssim40$ \textmu m.

%The primary purpose of dispersant application following an oil spill is the resulting reduction in droplet size, which increases the total amount of available oil-water interface. This fact underscores the importance of studying the effects of droplet size on the mechanics of oil biodegradation when assessing dispersant effectiveness, as was done here.

Our experimental design allows us to decouple the effects of droplet size and dispersant concentration on droplet colonization by keeping the droplet size distribution constant across experiments with and without dispersants. In bulk reactor experiments, the addition of dispersants necessarily leads to a reduction in droplet size. While this design reproduces some important features of real oil spill responses, it can lead to confounding results, and make an assessment of the actual effect of dispersants on cell growth and attachment impossible. Using our microfluidic device, experimental protocols were designed to allow for direct visualization of oil droplets and distinguish specific interactions between dispersants, oil, and bacteria in conditions that are realistic for oil-contaminated marine environments. In contrast to batch-reactor experiments, the constant flow through the chamber leads to rapid dilution of dispersants and of soluble components of oil. Additionally, droplets are prevented from interacting with one another by being pinned to the glass slide.
 
One notable difference between our experiments and real oil-spill scenarios is that in vivo, dispersants are often mixed with the oil before droplets form in the water column. For example, dispersants may be sprayed onto an oil slick, or injected directly into oil spilling from a wellhead. Nevertheless, since bacteria only attach to the droplet surface and do not penetrate into the oil, the absence of dispersants in the bulk of the oil droplets is not expected to affect our experimental results. Here, we are interested in determining if dispersants can enhance or hinder biodegradation purely by affecting cell attachment or growth at the droplet surface. Therefore the application of dispersants to droplets after droplet formation is appropriate.

The effect of dispersants on initial surface attachment of cells was investigated by measuring buckling times in experiments where Corexit 9500 decorates the droplet surface before bacterial colonization (protocol 1). In this case, the mean buckling time is about 1 h longer compared to the control, suggesting a slight decrease in initial cell attachment. The large variation in the buckling times at low radii remains significantly larger than the delay caused by the addition of dispersants. The initial concentration of Corexit (100 ppm) is consistent with values used in the literature to study degradation of oil with dispersants \cite{Dewangan2018,Fu2017,Gong2017}, as well as the toxicity of dispersants to various marine organisms \cite{Adeyemo2015,Wise2011}. The total volume of Corexit injected in protocol 1 corresponds to a DOR of 1:50. However, 100 ppm is much higher than the concentration of dispersants which persist in marine environments as oil is biodegraded: usually less than 1 ppm \cite{Kujawinski2011}. Thus, the rapid dilution of dispersants in protocol 1 creates a more realistic environment to study the colonization of oil droplets by bacteria.

We distinguish between the effects of dispersants on the surface attachment and growth of bacteria at the surface by varying when dispersants are added into the microfluidic device. In protocol 2, Corexit was injected after the cell suspension to quantify its effect on interfacial growth, which led to an increase in buckling times of about 2 h compared to the control. Because of the absence of dispersants dilution, protocol 2 is less representative of real oil-spill scenarios than protocol 1. Nevertheless, these experiments were needed to decouple the different potential effects of dispersants on the colonization of oil droplets by bacteria. Since, in protocol 2, initial attachment took place in the absence of dispersants, the delay of buckling could be attributed to a change in the cell growth rate. However, experiments at a flat oil-water interface conclusively showed that the rate of cell growth and division is unaffected by Corexit.

In conjunction, these results suggest that the increase in buckling times in the presence of Corexit is caused neither by a change in initial surface attachment nor a change in cell growth rate at the droplet surface. A decrease in the cell growth rate is ruled out by the direct measurements of cell growth at flat interface, and a decrease in cell attachment is inconsistent with the longer buckling times observed when dispersants were injected after initial attachment took place. Instead, we believe that by altering the surface energy of the oil-water interface, dispersants alter the complex interactions between the cell monolayer and the droplet surface which ultimately gives rise to droplet deformations. When viewed this way, the increase in buckling times is counter-intuitive. A priori, a decrease in surface tension is expected to make the surface easier to deform, meaning it should take less time for the pressure that arises from confined cell growth to induce buckling. Here, we observe the opposite trend, meaning further work is required to understand how active mechanics of a cell monolayer cause deformations of a liquid-liquid interface. 

Regardless of the precise mechanism at work, the absence of a decrease in cell attachment and growth rate suggests that, at the scale of individual bacteria encountering oil droplets, dispersants are unlikely to have a significant effect on biodegradation rates. Additionally, the delay of buckling is small compared to the variation in buckling times at $R\lesssim40$ \textmu m. Thus, when assessing the impact of Corexit on the biodegradation of immiscible oil by \textit{A. borkumensis}, one can consider the reduction in droplet size to be the dominant effect. Our Monte Carlo simulations of cell encounters and their subsequent colonization of droplets showed that the variation in buckling times is due to diffusive encounters and the stochastic nature of attachment. At very high cell concentrations, the encounter rate is expected to be large and the variation in attachment due to stochastic encounters is small compared to the large number of total attached cells. In this regime, small droplets are always colonized faster than large ones. At low cell concentrations, on the other hand, it can take a very long time for bacteria to encounter small droplets, meaning that many large droplets are actually colonized faster than smaller ones. 

These insights from simulations suggest that the use of Corexit is likely to inhibit droplet colonization in environments with low cell concentrations by reducing the mean droplet size. The low encounter rate could lead to oil being degraded more slowly. On the other hand, the use of Corexit is likely to be effective in enhancing biodegradation when the cell concentration is sufficiently high for small droplets to encounter cells quickly. In most marine environments, the background concentration of oil-degrading bacteria like \textit{A. borkumensis} is very low. However, following an oil spill, they can proliferate rapidly and make up a large portion of the total bacterial population \cite{Head2006}. Our results show that it is important to consider these population dynamics when designing oil spill responses involving the use of chemical dispersants.

\section{Conclusion}

We conducted experiments using a microfluidic device to examine the dynamics of oil droplet colonization by marine bacteria which leads to buckling of the droplet surface. Agent-based simulations were used to confirm and better understand the dependence of buckling times on droplet size observed in experiments. The time it takes for buckling to take place is a measure of how quickly cells attach to and grow on the surface. Buckling time was found to increase with increasing droplet size, and increased slightly whith the chemical dispersant Corexit 9500. Notably, the variation in buckling times was much larger for small droplets, suggesting that a reduction in droplet size due to dispersant use could have the unintended effect of reducing biodegradation rates. Corexit delayed buckling regardless of whether it was added before or after the cell suspension, so the effect was not due to a decrease in the initial attachment of cells to droplets. Additionally, direct observation of cell growth also showed that the doubling time was unaffected by the presence of Corexit. Further study is required to understand the precise mechanism of buckling and how dispersants can impede interfacial deformations despite reducing surface tension.

This study provides direct observation of oil droplets being colonized by oil-degrading bacteria with unprecedented spatial and temporal resolution. By varying the order in which dispersants and bacteria come into contact with the oil, new light was shed on the complex interfacial interactions between bacteria and chemical dispersants. This experimental approach can thus resolve apparent contradictions in the existing literature, and can easily be extended to study other oils, bacteria, and dispersants which play a role in oil spill remediation. 

\section{Acknowledgments}

This material is based on work supported by the Oil Spill Recovery Institute under Contract No. 20-10-06. 

%------------------%
%%% bibliography %%%
%------------------%
% \begingroup
%\setstretch{1.0}
\bibliography{refs2}
% \endgroup

\end{document}